\documentclass[10pt]{article}

\usepackage{microtype}
\usepackage{graphicx}
\usepackage{booktabs} %
\usepackage{xspace}
\usepackage{soul}
\usepackage{caption}
\usepackage{subcaption}
\usepackage{xcolor}
\usepackage{hyperref}
\usepackage[most]{tcolorbox}  %
\usepackage{enumitem}  %
\usepackage{tikz}
\DeclareRobustCommand\numcircledtikz[1]{\tikz[baseline=(char.base)]{
    \node[shape=circle,draw,fill,inner sep=1pt] (char)
    {\textcolor{white}{#1}};}}
\usepackage{array, makecell}

\usepackage{float}  %
\usepackage{placeins}

\usepackage{lipsum}
\usepackage{eso-pic}
\usepackage{xparse}
\newlength{\badgewidth}
\newlength{\badgegap}
\newcommand{\badgeList}{}

\NewDocumentCommand{\addTopRightBadge}{O{} m}{%
\gappto{\badgeList}{\href{#1}{\includegraphics[width=\badgewidth]{#2}}\hspace{\badgegap}}%
}

\newcommand{\placeTopRightBadges}{%
\AddToShipoutPictureBG*{%
\put(\LenToUnit{\paperwidth - 0.25cm - \badgewidth},\LenToUnit{\paperheight - 2.25cm}){%
\makebox[0pt][r]{\badgeList}%
}%
}%
}

\addTopRightBadge{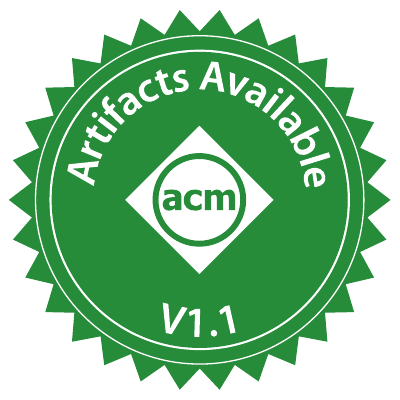}
\addTopRightBadge{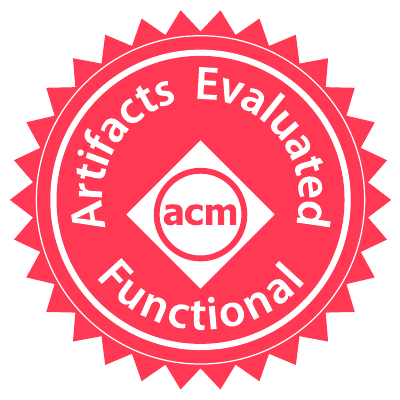}

\placeTopRightBadges

\newcommand{\squishlist}{
   \begin{list}{$\bullet$}
    { \setlength{\itemsep}{0pt}      \setlength{\parsep}{3pt}
      \setlength{\topsep}{3pt}       \setlength{\partopsep}{0pt}
      \setlength{\leftmargin}{1.0em} \setlength{\labelwidth}{1em}
      \setlength{\labelsep}{0.5em} } }
\newcommand{\squishend}{
    \end{list}  }
    
\newcommand{\name}{Marconi\xspace}
\newcommand{\fulltitle}{Prefix Caching for the Era of Hybrid LLMs}

\usepackage[accepted]{mlsys2024}

\mlsystitlerunning{\name: \fulltitle}

\begin{document}

\twocolumn[
\mlsystitle{\name: \fulltitle}

\mlsyssetsymbol{intern_aws}{*}

\begin{mlsysauthorlist}
\mlsysauthor{Rui Pan}{princeton,intern_aws}
\mlsysauthor{Zhuang Wang}{aws}
\mlsysauthor{Zhen Jia}{aws}
\mlsysauthor{Can Karakus}{aws}
\mlsysauthor{Luca Zancato}{aws}
\mlsysauthor{Tri Dao}{princeton}
\mlsysauthor{Yida Wang}{aws}
\mlsysauthor{Ravi Netravali}{princeton}
\end{mlsysauthorlist}

\mlsysaffiliation{princeton}{Princeton University}
\mlsysaffiliation{aws}{AWS}

\mlsyscorrespondingauthor{Rui Pan}{ruipan@princeton.edu}

\mlsyskeywords{Machine Learning, MLSys}

\vskip 0.3in

\begin{abstract}
Hybrid models that combine the language modeling capabilities of Attention layers with the efficiency of Recurrent layers (e.g., State Space Models) have gained traction in practically supporting long contexts in Large Language Model serving. Yet, the unique properties of these models complicate the usage of complementary efficiency optimizations such as prefix caching that skip redundant computations across requests. Most notably, their use of in-place state updates for recurrent layers precludes rolling back cache entries for partial sequence overlaps, and instead mandates only exact-match cache hits; the effect is a deluge of (large) cache entries per sequence, most of which yield minimal reuse opportunities. We present \name{}, the first system that supports efficient prefix caching with Hybrid LLMs. Key to \name{} are its novel admission and eviction policies that more judiciously assess potential cache entries based not only on recency, but also on (1) forecasts of their reuse likelihood across a taxonomy of different hit scenarios, and (2) the compute savings that hits deliver relative to memory footprints. Across diverse workloads and Hybrid models, \name{} achieves up to 34.4$\times$ higher token hit rates (71.1\% or 617 ms lower TTFT) compared to state-of-the-art prefix caching systems.

\end{abstract}
]

\printAffiliationsAndNotice{\textsuperscript{*}Work done during internship at AWS.}  %

\section{Introduction}
\label{s:intro}

\begin{figure*}[t]
    \centering
    \begin{subfigure}[c]{0.32\linewidth}
        \centering
        \includegraphics[width=\linewidth]{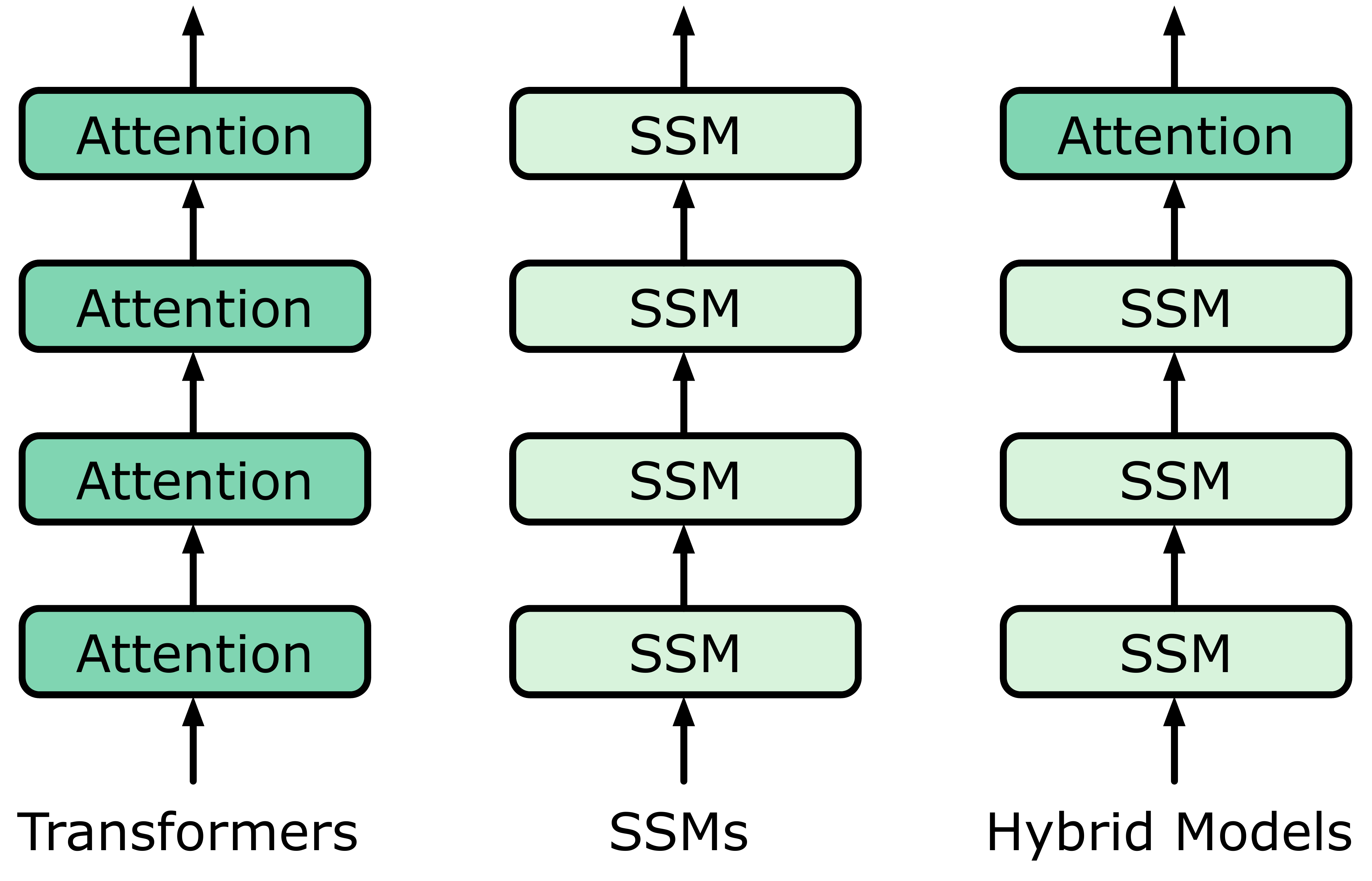}
        \caption{Different model architectures. Hybrid models usually mix in one Attention layer for every 6-10 SSM layers.}
        \label{fig:overview_model_arch}
    \end{subfigure}
    \hfill
    \begin{subfigure}[c]{0.32\linewidth}
        \centering
        \includegraphics[width=\linewidth]{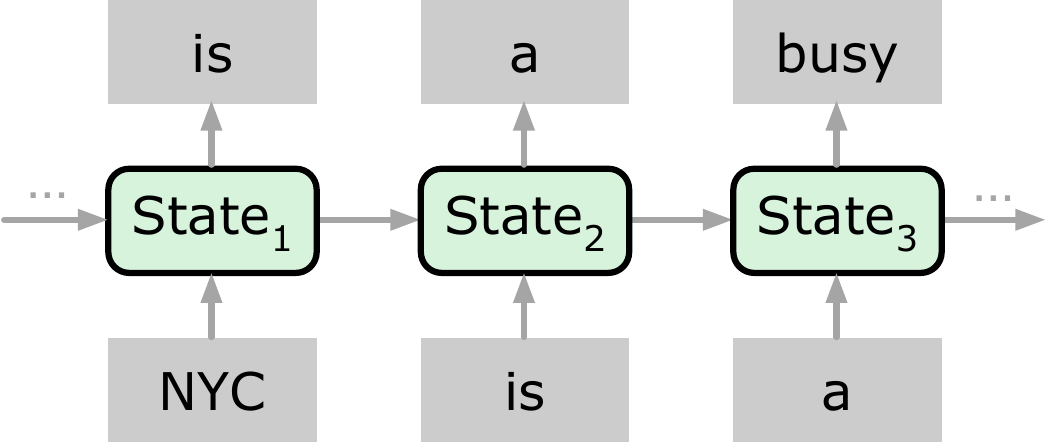}  %
        \caption{SSMs update their states recurrently and in place.}
        \label{fig:overview_ssm}
    \end{subfigure}
    \hfill
    \begin{subfigure}[c]{0.32\linewidth}
        \centering
        \begin{tabular}{|c|c|c|}
            \hline
             & Attention & SSM \\
            \hline
            \makecell{Computational \\ Complexity} & $O(L^2)$ & $O(L)$ \\
            \hline
            \makecell{Inference-Time \\ Memory} & $O(L)$ & $O(1)$ \\
            \hline
        \end{tabular}
        \caption{Prefill complexity comparison between quadratic Attention layers and subquadratic SSM layers.}
        \label{fig:overview_complexity}
    \end{subfigure}
    \vspace{-12pt}
    \caption{Overview of Hybrid models.}
    \vspace{-5pt}
    \label{fig:overview}
\end{figure*}

\begin{figure}[t]
    \centering
    \includegraphics[width=0.9\columnwidth]{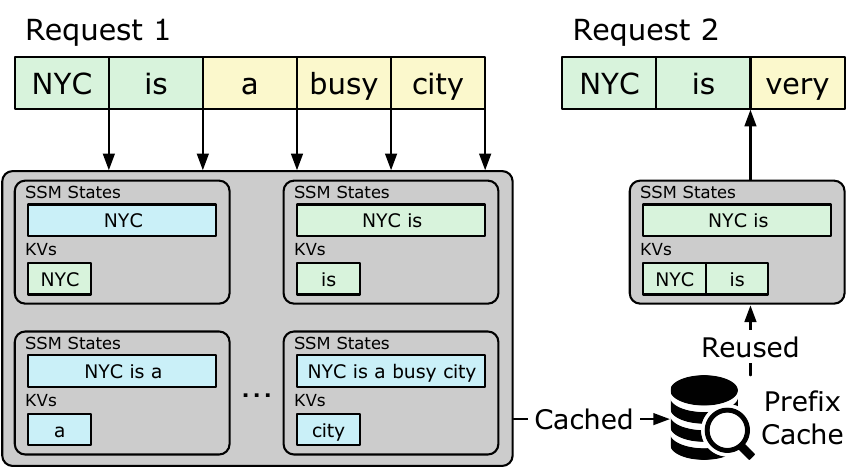}
    \vspace{-8pt}
    \caption{Prefix caching reuses model states of common prefixes (green) across requests, accelerating inference. Fine-grained checkpointing results in many sparsely-hit entries (blue).}
    \label{fig:prefix_caching}
    \vspace{-10pt}
\end{figure}

The emergence of large language models (LLMs) has empowered many applications including chatbots~\cite{chatgpt}, AI-powered search~\cite{perplexity}, coding assistants~\cite{github_copilot}, and more. Owing to increasing workload complexity and the evolution of inference-time enhancements such as few-shot prompting~\cite{fewshot_prompting} and chain-of-thoughts~\cite{cot}, recent years have witnessed a push towards \emph{longer context windows} during serving. Indeed, the accuracy improvements from multi-step reasoning~\cite{dspy, react}, detailed prompt templates~\cite{prompt_engineering_survey}, and increased samples in few-shot prompting~\cite{fewshot_prompting} all require expanded context sizes.

To keep pace with these trends, traditional foundation models like Transformers have been extended to support longer context windows, e.g., Gemini 1.5~\cite{gemini} and Claude 3~\cite{claude3} both have 1M window size. However, their intrinsic reliance on the Attention mechanism harms long-context serving efficiency, particularly due to its quadratic compute complexity and large memory footprint for housing KV cache states~\cite{loongserve,infinigen}. 
As a result, new recurrent and subquadratic model architectures like State Space Models (SSMs)~\cite{mamba,mamba2} have emerged to offer lower compute and memory costs for long-context serving (Fig.~\ref{fig:overview_complexity}), albeit with mild accuracy degradations~\cite{mamba_empicircal_study}. 
These subquadratic layers are commonly mixed with several full Attention layers to produce \emph{Hybrid LLMs} (Fig.~\ref{fig:overview_model_arch}) that lower serving costs while maintaining Attention's superior recall and in-context learning capabilities~\cite{mamba_empicircal_study,jamba,jamba1.5,zamba}.

Although Hybrid LLMs improve per-request efficiency, they complicate the cross-request efficiency wins that prior efforts have shown to be especially effective in long-context scenarios, i.e., \emph{prefix caching} that reuses model states for common prefixes across requests, thereby skipping computation without harming accuracy~\cite{vllm, sglang, cachedattention, infercept}.
The primary challenge is that SSM model states are updated in place (Fig.~\ref{fig:overview_ssm}), so states at the end of a sequence cannot be rolled back to represent a prefix of the sequence. This presents a dilemma for serving systems. On one hand, maximizing reuse opportunities for future (arbitrary) workloads mandates caching fine-grained state checkpoints at regular intervals, e.g., every 256 tokens. On the other hand, increased checkpointing frequency inflates the number of cache entries generated per sequence, each of which is \emph{large} (due to the sheer size of SSM states) but most of which present limited reuse opportunities, i.e., \emph{sparsely-hit}. The net effect is cache thrashing with low-utility entries (Fig.~\ref{fig:prefix_caching}), and a poor memory-vs-compute savings tradeoff (\S\ref{s:challenges}).

We present \textbf{\emph{\name{}}},\footnote{``Marconi plays the mamba, listen to the radio, don't you remember?'' -- Lyrics of \textit{We Built This City}, song by Starship} the first prefix caching system designed to support Hybrid LLMs.
For a given sequence, \name{} simultaneously manages cache entries for both SSM states and KVs, ensuring that \emph{all} of the preceding sequence states for each cached entry are also present to support reuse. Moreover, \name{} eschews traditional caching logic based on recency~\cite{vllm, sglang, preble}, and instead introduces novel admission and eviction strategies that value potential cache entries with awareness of the aforementioned SSM overheads; we describe these in turn.

To cope with the large size and ``all or nothing'' nature of SSM cache entries (i.e., hits only arise on exact sequence matches, unlike with KVs), \name{} adopts a judicious admission strategy, whereby only SSM candidates with high reuse likelihood are accepted. Our driving insight is that, despite lacking knowledge of upcoming requests, the reuse potential of each SSM state can be sufficiently estimated when assessed against the taxonomy of potential prefix reuse scenarios. Specifically, token redundancy arises across requests as either (a) purely-input shared prefixes, e.g., system prompts, or (b) a combination of input and output token sharing, e.g., conversation history. The former case typically arises across many requests, enabling \name{} to vet each candidate against previously-observed sequences. In contrast, for the latter case, \name{} assigns value only to SSM states that represent the last decoded token (which conversations typically resume from, as opposed to intermediate branching). We efficiently encapsulate this bookkeeping in a radix tree that highlights sequence overlap, and introduce a speculative insertion step prior to each prefill to determine the potential reuse opportunties for the upcoming sequence (and the resultant checkpointing required).

For eviction, our main observation is that, with Hybrid LLMs, when assigning value to the cache entry for a given sequence, we have to account for both KVs and SSM states which exhibit different tradeoffs between memory and compute savings. Specifically, whereas the size of KVs' entries for a sequence is (linearly) proportional to the compute savings from reusing that sequence, SSM state sizes are fixed and unrelated to sequence length and compute savings. To enable more holistic management of cache entries, \name{} introduces a new FLOP-aware eviction policy that assesses candidates for eviction based not only on recency/popularity, but also the potential compute savings they deliver (normalized against the space they consume in the cache). This inherently trades the hit rate of shorter sequences for longer ones -- an especially desirable tradeoff given the superior efficiency of Hybrid models over Transformers.

We evaluated \name{} on a wide range of workloads, request arrival patterns, cache sizes, and Hybrid model architectures. Overall, we find that \name{} improves cache hit rates by an average of 4.5-34.4$\times$ compared to state-of-the-art caching systems (e.g., vLLM~\cite{vllm}, SGLang~\cite{sglang}) that are extended to support Hybrid models.
The win in token hit rate translates to latency savings of 36.1-71.1\% (103.3-617.0ms) for P95 TTFT.
Microbenchmarks show that \name{} performs better in scenarios with longer contexts, higher ratios of SSM layers, and larger SSM state dimensions — trends that align with recent model developments. Marconi is open sourced at \url{https://github.com/ruipeterpan/marconi}.

\section{Background}
\label{s:background}

In this section, we provide background on the efficient recurrent architectures (exemplified by SSMs), Hybrid models that contain these alternative architectures, and prefix caching for efficient LLM inference.

\subsection{State Space Models and Hybrid Models}

Token generation in LLM inference involves two main phases. First, the prefill phase processes an input sequence, generating the model's internal states (e.g., KV cache/KVs in Attention layers) for each layer of an LLM, and outputs the first new token. The decoding phase then utilizes the internal states to perform autoregressive token generation.

During prefill, generating the first token depends on all previous tokens. In Transformers, the self-attention mechanism calculates how each token in the sequence ``attends'' to every other token.  Consequently, the Attention mechanism incurs quadratic computational complexity~\cite{flashattention}, which quickly bottlenecks GPU compute as sequence lengths scale~\cite{sarathi}.
Furthermore, the size of the KVs in Attention layers grows linearly with sequence length, resulting in a large inference-time memory requirement~\cite{loongserve,infinigen}.

State space models (SSMs) and more generally linear RNNs and linear attention, such as Mamba~\cite{mamba, mamba2}, address these inefficiencies by selectively
``compressing'' previous context into a recurrent and compact representation.
The recurrent representation is used alongside the previous token to update the recurrent representation in place, as shown in Fig.~\ref{fig:overview_ssm}.
Because the SSM states maintain a constant size, memory consumption remains fixed regardless of sequence length, and the computational complexity scales linearly, rather than quadratically, with the sequence length (Fig.~\ref{fig:overview_complexity}).
Although pure SSM models outperform Transformers on many NLP tasks, they lag behind on certain workloads that require strong recall or in-context learning capabilities~\cite{mamba_empicircal_study,needleinahaystack}. 
To balance inference efficiency and model capability, SSM-Attention \textit{Hybrid models} (Fig.~\ref{fig:overview_model_arch}) have been proposed. These models blend quadratic Attention and subquadratic SSM layers, typically interleaved in a specific ratio (commonly 1 Attention layer for every 6-10 SSM layers~\cite{mamba_empicircal_study,jamba,zamba}). When compared to Transformers of equivalent scale trained on the same datasets, Hybrid models demonstrate superior performance across a wide range of tasks while preserving most of the efficiency advantages of SSM layers (up to 8$\times$ faster)~\cite{mamba_empicircal_study}. Many Hybrid models have been productionized~\cite{cartesia_rene,jamba,jamba1.5,zamba,zamba2_7b,zamba2_mini,mamba_empicircal_study,samba}, with the largest being Jamba 1.5 at 398B parameters~\cite{jamba1.5}.

\subsection{Prefix Caching}

Shared prefixes across requests, including input tokens and sometimes output tokens, are common across many LLM applications.
For example, question-answering workloads often share a detailed system prompt combined with few-shot examples that provide instructions and demonstrations~\cite{react}. Similarly, coding agents interact with the environment in multiple rounds, where each new request consists of a trajectory of past environment interactions and new observations and actions~\cite{sweagent,opendevin}. Redoing the prefill of these shared prefixes for all requests leads to many redundant computations, hurting both throughput and latency. Prefix caching mitigates this by caching and reusing the model’s internal states that represent the common prefixes (Fig.~\ref{fig:prefix_caching}), achieving a lower time to first token (TTFT) latency, lower tail time per token (TPT) latency\footnote{Even though prefix caching is a prefill-only optimization, a lower prefill latency also reduces the tail TPT for high-throughput LLM inference engines~\cite{sarathi}.}, and higher prefill throughput (measured in tokens/s).  %

Many research and production systems have been proposed to reap the benefits of prefix caching in Transformer inference~\cite{vllm,sglang,cachedattention,preble,infercept,claude,characeterai,mooncake,openai_prefix_caching}. On startup, these systems provision blocks of GPU/CPU memory for caching the prefix states. 
Before prefilling a new sequence, the inference engine looks up the prefix cache for the longest matching prefix. Upon a cache hit, the corresponding prefix is fetched before prefilling. After decoding the final token of the sequence, to favor recency, the system admits the model states of all tokens of the new sequence into the cache. To reduce memory fragmentation, the KVs are usually partitioned into fixed-sized token blocks in the prefix cache, each housing the KVs of $x$ tokens where $x$ is the block size~\cite{vllm}, and existing token blocks are evicted to make room if the cache is full. Because KVs have a sequence dimension, managing and evicting model states is efficient and flexible. E.g., if we have the KVs of a sequence of tokens $1...q$ and want to retain the KVs of the prefix $1...p$ ($p<q$), tensor slicing can be performed on the sequence dimension of the full KVs.  %

\section{Challenges of Prefix Caching with Hybrid LLMs}
\label{s:challenges}

Compared to Attention, SSM's properties greatly benefit its per-request computational complexity and memory consumption. However, the very properties that make SSMs more efficient also complicate prefix caching, an important optimization for cross-request efficiency wins.
The key reason lies in how SSM updates its internal states:
SSMs recurrently and efficiently compress the history of tokens into a compact state and update the recurrent state in place by overwriting the previous states, keeping the memory usage constant.
Further, while SSM states are generally smaller than the KVs of \textit{whole sequences} in Attention, they aim to capture the same cross-token information as KVs do and thus are typically 10-100x larger than a \textit{single token}'s KVs (Appendix~\ref{s:appendix}).
In summary, SSM states exhibit the following key properties:

\begin{tcolorbox}[colback=green!2!white,colframe=green!35!black,title=SSM State Properties]
\begin{enumerate}[noitemsep,topsep=0pt,parsep=0pt,partopsep=0pt]
    \item SSM states are constant-sized regardless of how many tokens they represent.
    \item SSM states are updated in place, so a sequence's states cannot be rolled back to represent its prefixes.  %
    \item SSM states are orders of magnitude larger than the KVs of a single token.
\end{enumerate}
\end{tcolorbox}

\begin{figure}[t]
    \centering
    \begin{subfigure}[t]{0.44\linewidth}
        \centering
        \includegraphics[width=\linewidth]{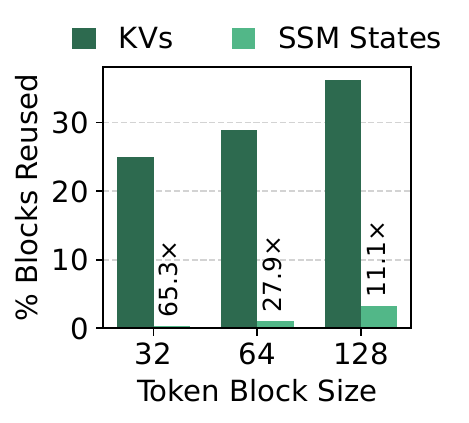}
        \caption{Token block reuse rate comparison between KVs and SSM states.}
        \label{fig:cache_usage_breakdown}
    \end{subfigure}
    \hfill
    \begin{subfigure}[t]{0.54\linewidth}
        \centering
        \includegraphics[width=\linewidth]{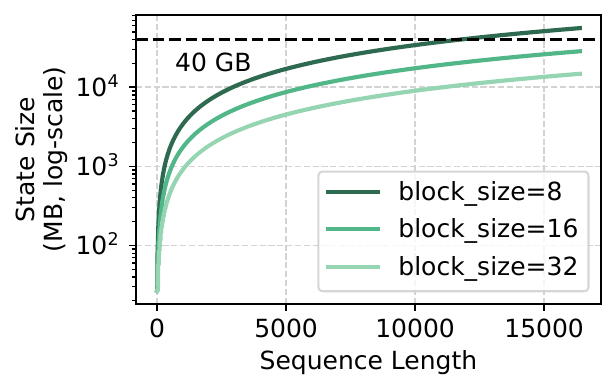}  %
        \caption{Total size of a Hybrid model's cache entries as sequence length scales.}
        \label{fig:periodic_caching}
    \end{subfigure}
    \caption{Fine-grained caching of token blocks results in many SSM states being cached. This creates sparsely-hit entries in which many SSM states are never reused (a), underutilizing the precious cache capacity. Worse, this creates a huge memory usage even for a single sequence of a 7B model (b), overwhelming and thrashing the cache.}
    \label{fig:motivation}
    \vspace{-5pt}
\end{figure}

The core challenge of prefix caching in Hybrid LLM inference is that SSM states exhibit ``all or nothing'' reusability: To realize prefix reusing, we need all prefix tokens' KVs for each Attention layer and one SSM state that exactly matches all prefix tokens for each SSM layer.
However, if future requests only use a prefix of a sequence, such as tokens $1...p$ from a sequence $1...q$ ($p<q$), the SSM layers cannot reuse states that represent $1...q$ (property \numcircledtikz{2}) and neither can the KVs be reused by Attention layers, as prefix reusing is bottlenecked by the layer with the least reusing opportunities. %
To maximize reuse opportunities for future requests that might reuse arbitrary prefixes of the most recent request, fine-grained checkpointing (every $x$ tokens) of many previously overwritten SSM states is required. This creates an equal-sized token block for each interval, containing KVs for $x$ tokens and SSM states that represent all prior tokens (Fig.~\ref{fig:prefix_caching}).
However, this elicits two challenges for existing prefix caching systems. To demonstrate this, we analyze one of the experiments in~\S\ref{s:evaluation}. No existing systems support prefix caching for Hybrid models, so we extend vLLM~\cite{vllm} using its caching policy to support Hybrid models.

\textbf{Cache underutilization.}
While reusing KVs requires accessing all prior tokens' KVs, reusing SSM states only needs the last token block. \ul{This results in sparsely-hit cache entries on a sequence level, with many low-utility SSM states never accessed after admission.} Fig.~\ref{fig:cache_usage_breakdown} shows that with a block size of 32, 25.0\% of the token blocks' KVs are reused by future requests, but a mere 0.4\% of SSM states are reused, a 65.3$\times$ difference. Using larger blocks mitigates both issues by reducing the checkpointing granularity and the number of SSM states checkpointed. However, the issues persist (3.3\% reuse rate for block size 128 which exceeds what vLLM natively supports), and larger block sizes lead to internal memory fragmentation of KVs within token blocks, impacting memory utilization~\cite{vllm}.

\textbf{High memory usage.} Worse, \ul{fine-grained checkpointing quickly leads to an excessive memory usage due to the SSM states' size.}
For example, with block size 16, the state size of an SSM layer is 4$\times$\footnote{\texttt{d\_state} $/ (2\cdot$ \texttt{block\_size}$)$. The 2 stands for the key and value tensors in KVs. See Appendix~\ref{s:appendix} for more details.} larger than the KVs of an Attention layer in the token block because of property \numcircledtikz{3}. Since there are many more SSM layers than Attention layers in Hybrid models for efficiency~\cite{mamba_empicircal_study}, their states quickly overwhelm the GPU HBM (and possibly CPU RAM): for a 7B model, a single sequence of 10K tokens consumes 17.4 GB (Fig.~\ref{fig:periodic_caching}), 3.3$\times$ bigger than a Transformer of the same size. 
Even if sequences are short and the states are evicted soon after a request finishes inference, they must still be admitted into the prefix cache, potentially requiring eviction of entries with higher utility, leading to frequent cache thrashing and low cache hit rate.

In conclusion, existing prefix caching systems face a dilemma in Hybrid LLM inference: maximizing reuse opportunities mandates fine-grained state checkpointing but doing so creates large yet sparsely-hit entries that overwhelm and thrash the limited cache capacity. 
Therefore, a judicious cache management scheme is needed to better reap the benefits of prefix caching in Hybrid LLM inference.

\section{Design and Implementation}
\label{s:design}

\begin{figure}[t]
    \centering
    \includegraphics[width=0.99\columnwidth]{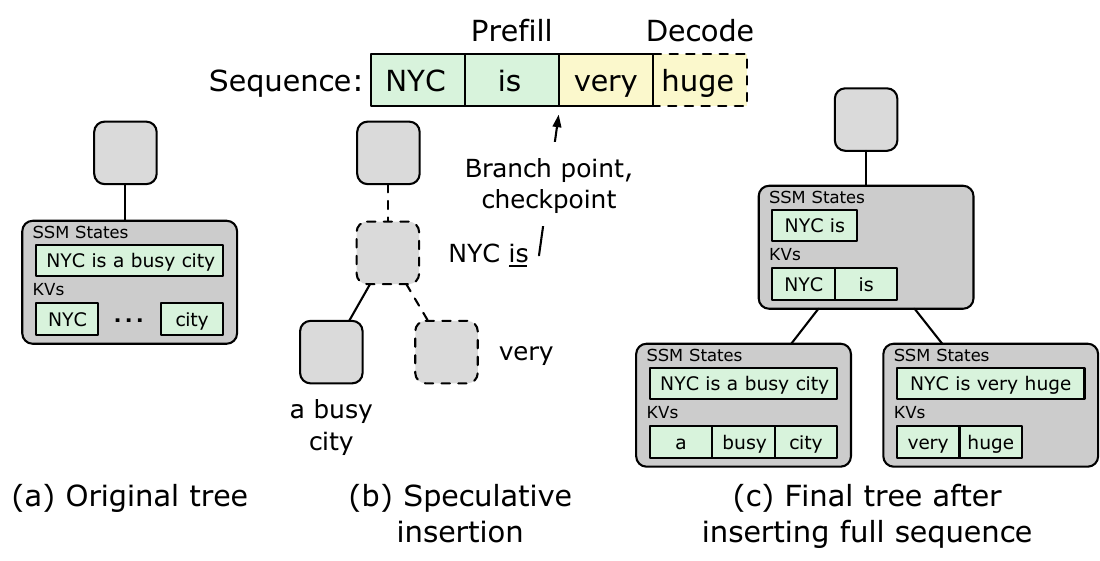}
    \caption{\name{} performs a speculative insertion to check if inserting the prefill segment of a sequence results in an intermediate node. If so, the SSM states at the branch point are checkpointed. States at the last decoded token are checkpointed in any case. For ease of visualization, we associate model states with nodes rather than edges.}  %
    \label{fig:spec_insertion}
    \vspace{-5pt}
\end{figure}

\name{} is the first prefix caching system designed to accommodate the unique characteristics of Hybrid models.
It is designed to support models with arbitrary layer compositions, including Hybrid models, pure Transformers, and pure SSMs.
Its primary goal is maximizing the cache utilization to reduce redundant computation and minimize TTFT latency.
To improve cache utilization and prevent the large SSM states from thrashing the cache, \name{} judiciously admits SSM states (\S\ref{ss:admission}), only accepting states with a high reuse likelihood based on a taxonomy of potential prefix reuse scenarios. The bookkeeping of requests is realized via a radix tree that represents past request overlap.
Once admitted, as states of all layers need to work in conjunction and represent the same prefix tokens to realize prefix caching, \name{} opts to manage different types of model states holistically in the same tree (Fig.~\ref{fig:spec_insertion}), where each node contains the SSM states and KVs of a sequence, rather than disaggregating the cache space for different types of layer states. %
For eviction (\S\ref{ss:eviction}), to account for the different tradeoffs between memory and compute savings of different layers' states, \name{} introduces a FLOP-aware eviction policy that balances recency and potential compute savings delivered by reusing entries.
We describe the implementation details in \S\ref{ss:implementation}.

\subsection{Cache Admission}
\label{ss:admission}

\name{} aims to cache the states with high reuse likelihood during admission.
Although prefix reuse patterns of future requests cannot be perfectly predicted, our key insight is that the reuse potential can be sufficiently estimated through a taxonomy of potential prefix reusing scenarios. Through extensive analysis of prefix reusing patterns in various real-world datasets and request traces~\cite{mooncake,characeterai,sglang} that represent traffic of both dedicated inference deployments~\cite{swebench} and public-facing APIs~\cite{chatgpt,sharegpt}, we classify the token composition of all reused prefixes into two types:

\begin{enumerate}[noitemsep,topsep=0pt,parsep=0pt,partopsep=0pt]
    \item \textbf{Purely input:} The prefix is a part of the input sequence from a previous input, such as system prompts~\cite{chatgpt}, instructions~\cite{react}, few-shot examples~\cite{mmlu}, self-consistency~\cite{self_consistency}, long-document QA~\cite{loogle}, etc.
    \item \textbf{Input and output:} The prefix consists of previous input and output tokens, such as conversation history in chatbots~\cite{cachedattention,characeterai,chatgpt}, past environment interactions for LLM agents~\cite{sweagent,react}, etc.
\end{enumerate}

This taxonomy allows \name{} to devise different mechanisms for identifying and caching the corresponding states. For the purely-input case, as the same prefix is usually shared across many requests, \name{} can observe and compare previous requests to identify these hot common prefixes. For the input-and-output case, \name{} only values SSM states that represent the last decoded token between conversation rounds, which conversations typically append to, as opposed to branch off from.

Leveraging the above insights, to estimate the reuse likelihood, \name{} uses a radix tree to bookkeep the request history, identify common prefixes, and map sequences to their states.
A radix tree is a space-efficient prefix tree with edges labeled by sequences of varying lengths.
Within the tree, each edge is associated with the KVs of tokens it represents and the SSM states that represent all tokens prior to the last token in the edge (Fig.~\ref{fig:spec_insertion}c).
Nodes represent branch-off points in sequences: those with multiple children represent prefixes that are ``purely input'', whereas nodes with $\leq1$ child may represent ``input and output'' prefixes of future requests.
Because nodes represent high-utility states, \name{} caches states judiciously by only admitting SSM states represented by the last token of edges.
For input-and-output cases, \name{} simply checkpoints the state after the last decoding step.
Because the last token's KVs include all prior tokens, all KVs of the whole sequence are still effectively cached, which is the same as existing systems.
To identify ``purely input'' prefixes, prior to prefilling each sequence, \name{} employs a speculative insertion of the input tokens to see if new intermediate nodes will be created (Fig.~\ref{fig:spec_insertion}). If so, \name{} caches the prefix's states during prefill.

\textbf{Obtaining states during prefill.} 
After prefilling KVs in Attention, they can be partitioned and trimmed to represent subsequences. In contrast, SSM states cannot be rolled back to represent a prefix, so \name{} needs a new mechanism previously unnecessary for Transformers inference.
\name{} supports two main methods for checkpointing SSM states during prefill.
Some SSMs~\cite{mamba2,retnet,gla} perform chunked state passing during prefill, where the input sequence is split into chunks to compute intra-chunk states. 
For these models, we simply materialize and cache the state of the second-to-last chunk in the prefix. For example, when prefilling a sequence of 100 tokens using chunk size 32, if we need to cache the state at token 80, we can checkpoint the state at token 64. This approach may miss some prefix caching opportunities within a chunk but introduces minimal runtime overhead. 
Optionally, custom kernels can be developed to quickly roll the state forward by a few tokens to reach the exact location.
For models that don't support chunked state passing~\cite{mamba,jamba}, \name{} performs a two-pass prefill to get the precise state at the prefix. For example, the first pass prefills the first 80 tokens to generate the prefix state, while the second pass starts from the prefix state and prefills the remaining 20 tokens.

\textbf{Tradeoffs.}
Compared to fine-grained checkpointing, \name{}'s judicious admission has slight drawbacks, but its tremendous benefits compensate for the drawbacks.
Because the states of the last decoded token are immediately cached for all sequences, ``input and output'' prefixes can be reused instantaneously. In contrast, ``purely input'' prefixes only benefit from reusing starting from the third occurrence of the prefix, as \name{} uses the second occurrence to identify the prefix and checkpoint its states.
While this approach sacrifices the benefit of reusing a ``purely input'' prefix on its second occurrence, these prefixes are typically shared across many requests. As a result, missing savings on a single request has a negligible impact on overall savings.
On the other hand, judicious admission reduces coverage and slightly limits the potential reusability of arbitrary prefixes, as only up to two SSM states are admitted per sequence. However, due to the huge number of low-utility SSM states rejected from admission by \name{}, this altruistic approach significantly reduces the size and improves the utility of cached Hybrid model states, enhancing overall cache utilization.

\textbf{Comparison with SGLang.}
SGLang~\cite{sglang} is an existing prefix caching system for Transformers that also leverages a radix tree for mapping sequences to tokens and their KVs.
Different from SGLang, \name{} uses the past requests in the radix tree to determine which SSM states to cache by performing speculative insertions before prefilling an upcoming sequence.  %
Further, \name{}'s philosophy of judicious state admission fundamentally differs from SGLang, which admits the states of all tokens during admission and is no different from other prefix caching systems like vLLM~\cite{vllm}. We show how these differences yield superior cache efficiency in~\S\ref{s:evaluation}. 
Our system doesn’t invent new data structures for KV cache management, and instead builds on the rich prior work in this space that also aims to manage prefixes’ model states efficiently. Our main contribution lies in being the first system to redesign caching policies and their interactions with these data structures to practically address the unique properties of emerging Hybrid models.

\subsection{Cache Eviction}
\label{ss:eviction}

For eviction, our main observation is that KVs and SSMs in Hybrid model states exhibit different tradeoffs between memory and compute savings.
Specifically, whereas the size of KVs for a sequence is linearly proportional to the sequence length and (approximately) the compute savings from reusing that sequence, SSM state sizes are fixed regardless of sequence length and compute savings.
To quantify this difference, we propose a new metric, \textbf{FLOP efficiency}, to measure the compute savings (measured in the number of floating operations) per unit of memory achieved by reusing a prefix cache entry:
\vspace{-10pt}
\begin{equation}
    \mathit{flop\_efficiency} = \frac{\mathit{Total}\ \mathit{FLOPs}\ \mathit{across}\ \mathit{layers}}{\mathit{Memory}\ \mathit{consumption}\ \mathit{of}\ \mathit{all}\ \mathit{states}}
\end{equation}

\begin{figure}[t]
    \centering
    \includegraphics[width=0.8\columnwidth]{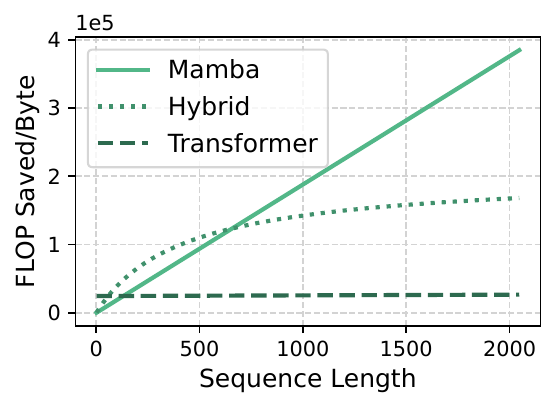}
    \caption{FLOP efficiency of the model states of different 7B models as the sequence length scales. The more SSM layers in the model, the steeper the increase in FLOP efficiency.}
    \label{fig:flops_eff_diff}
    \vspace{-5pt}
\end{figure}

Here, $\mathit{Total}\ \mathit{FLOPs}\ \mathit{across}\ \mathit{layers}$ denotes the sum of redundant compute across different types of layers (i.e., Attention, SSM, and MLP) circumvented by reusing the prefix entry, and $\mathit{Memory}\ \mathit{consumption}\ \mathit{of}\ \mathit{all}\ \mathit{states}$ denotes the total size of all stateful layers' states (i.e., Attention and SSM).
More details on FLOP efficiency are in Appendix~\ref{s:appendix}. Fig.~\ref{fig:flops_eff_diff} compares the FLOP efficiency of model states in three different 7B models: Transformers (Attention-only), Mamba (SSM-only), and Hybrid (with an Attention:SSM ratio of 1:6). 
The increase of FLOP efficiency as sequence length scales is steeper for models with a higher ratio of SSM layers as larger portions of the model states become more FLOP efficient.

Traditional prefix caching systems designed for Transformers don't need to consider FLOP efficiency because KVs' FLOP efficiency is near-constant, and most systems only use recency for eviction.
However, ignoring it in Hybrid LLM inference risks evicting states with higher FLOP efficiency but do not have the best recency.
To enable more holistic management of cache entries, \name{} introduces a \textbf{FLOP-aware eviction policy} that assesses candidates for eviction based not only on recency but also the potential compute savings they deliver (normalized against the space they consume in the cache). In addition to recency, \name{} accounts for the FLOP efficiency by computing a utility score $S$ that represents the utility for each radix node $n$:
\begin{equation}
    \mathit{S}(n) = \mathit{recency}(n) + \alpha \cdot \mathit{flop\_efficiency}(n)
\end{equation} 
This metric favors cache entries with higher recency, save more compute, and take less memory.
Both the \texttt{recency} and \texttt{flop\_efficiency} scores are normalized to the range $(0, 1)$ by comparing all nodes' last-accessed timestamps and FLOP saved/byte in the radix tree. 
During eviction, \name{} iteratively removes nodes with the lowest utility score until there is enough space to accommodate the new request's states. As such, child nodes' FLOP savings are calculated relative to parents' savings.

\textbf{Managing the balance.} \name{} manages the balance between favoring recency and FLOP efficiency by tuning $\alpha$.
A higher $\alpha$ emphasizes FLOP efficiency, while setting $\alpha$ to 0 falls back to LRU. 
\name{} manages the balance by observing the workload and retrospectively setting the best configuration.
On startup, \name{} sets $\alpha$ to 0 until the first eviction. Afterward, \name{} takes a snapshot of the radix tree and enters a bootstrap period, continuing to use LRU while bookkeeping token-level information of requests during the bootstrapping phase. The bootstrap period includes $5-15\times$ the number of requests seen before the first eviction, capturing a representative workload sample. Once sufficient requests have been observed, \name{} asynchronously launches a grid search over possible $\alpha$ values by replaying the bootstrap requests.
This grid search is parallelized across CPU cores, significantly speeding up the tuning process, typically taking just a few seconds, often shorter than the time required to prefill and decode a single request. After the grid search, \name{} adopts the $\alpha$ value that maximizes the hit rate.

\textbf{Comparisons with existing size-based eviction algorithms.} Cost-aware cache eviction for objects with variable sizes is a well-studied problem (e.g., GDSF~\cite{gdsf}). The KV size in Attention layers scales with sequence length, serving as a proxy for compute savings from cache hits, whereas SSM states are fixed-sized irrespective of sequence length or compute savings. Thus, size fails as a proxy in Hybrid LLM inference, where longer sequences (with greater compute savings) are represented by equally sized SSM states. We note that our system and techniques are complementary to such prior schemes; just as we combine FLOP efficiency with recency, our focus is to augment the estimators for cache entry importance used in prior work with a factor tailored specifically for Hybrid model reuse benefits.

\subsection{Implementation details}
\label{ss:implementation}

Alongside the custom admission and eviction policies, we made the following changes to accommodate Hybrid model states: (1) During eviction, all nodes with $\leq1$ children are considered for eviction, not just leaf nodes. The reasoning is that nodes with multiple children represent the common prefixes shared by multiple requests and should not be evicted
(unless they become stale, at which point all their children will be evicted first and the ex-parent nodes will become leafless nodes themselves, which make them subject for eviction), whereas intermediate nodes with a single child are unlikely to be reused more than once and still incur a memory cost for their SSM states. When an intermediate node is evicted, its SSM states are released, and its KVs are absorbed by its child node. (2) When a cache hit occurs, only the accessed node's timestamp is updated, unlike in existing systems where timestamps for all ancestor nodes are updated. In \name{}, previous SSM states are not reused (Fig.~\ref{fig:spec_insertion}c), and although ancestors' KVs are accessed, their KVs will be subsumed by child nodes if evicted. Thus, not updating ancestors' timestamps doesn't affect recency tracking.

\section{Evaluation}
\label{s:evaluation}

\begin{figure}[t]
    \centering
    \centering
    \includegraphics[width=\linewidth]{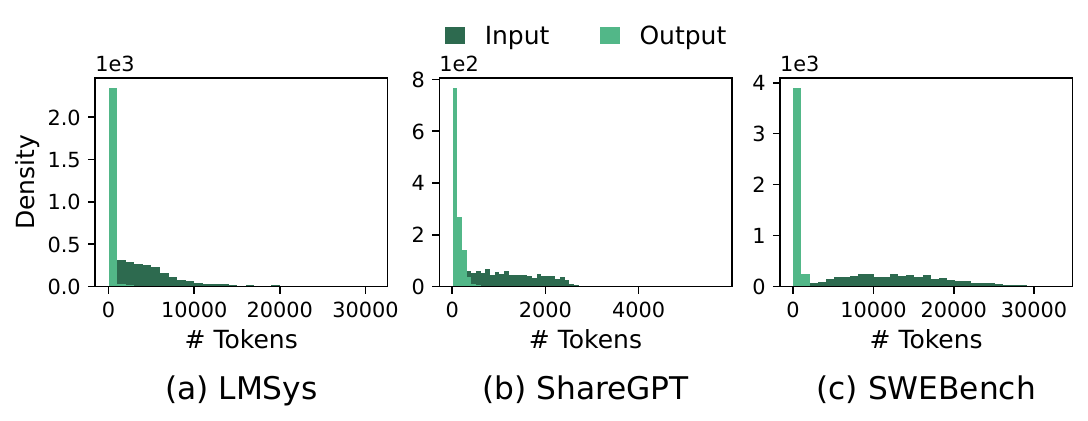}  %
    \vspace{-20pt}
    \caption{Input/output sequence length distributions per workload.}
    \vspace{-10pt}
    \label{fig:input_len_distribution}
\end{figure}

\begin{figure*}[t]
    \centering
    \begin{subfigure}{0.33\linewidth}
        \centering
        \includegraphics[width=\linewidth]{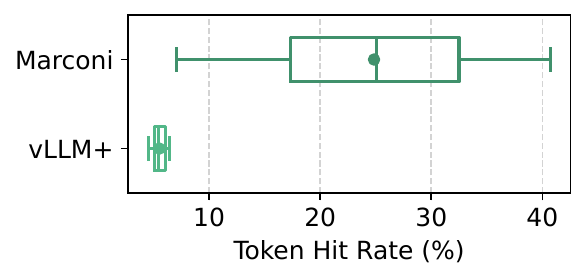}  %
        \caption{LMSys}  %
        \label{fig:xxx}
    \end{subfigure}
    \hfill
    \begin{subfigure}{0.33\linewidth}
        \centering
        \includegraphics[width=\linewidth]{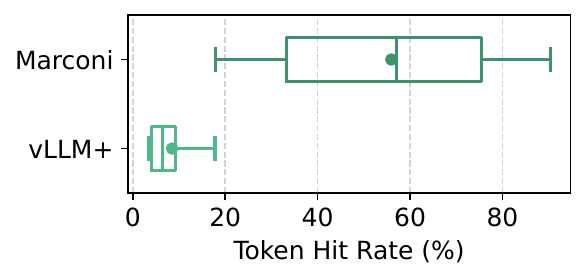}
        \caption{ShareGPT}  %
        \label{fig:xxx}
    \end{subfigure}
    \hfill
    \begin{subfigure}{0.33\linewidth}
        \centering
        \includegraphics[width=\linewidth]{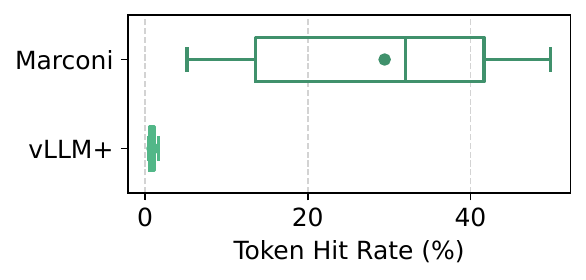}
        \caption{SWEBench}  %
        \label{fig:xxx}
    \end{subfigure}
    \vspace{-20pt}
    \caption{Comparison with vLLM+. With judicious cache admission, \name{} utilizes the limited cache space to retain states with higher utility, improving the token hit rate significantly over vLLM+.} %
    \label{fig:main_results}
    \vspace{-15pt}
\end{figure*}

We evaluated the performance of \name{} under various workloads with different datasets, request arrival patterns, cache sizes, and model architectures. Our key findings are:

\begin{enumerate}[noitemsep,topsep=0pt,parsep=0pt,partopsep=0pt]
    \item Overall, \name{} improves token hit rate by an average of 4.5-34.4$\times$ compared to fine-grained checkpointing, reducing the P95 TTFT by up to 71.1\% (617.0 ms) compared to baseline prefix caching systems.
    \item Compared to LRU, \name{}'s FLOP-aware eviction improves the token hit rate by 19.0-219.7\%. It achieves higher token hit rates and FLOP savings by trading off hit rate of shorter sequences to boost hit rate for longer sequences, a desirable tradeoff given the efficiency of Hybrid models over Transformers.
    \item \name{} performs better when sequences are long, the SSM layer ratio is high, and the SSM state dimension is large -- trends that align with recent model developments.
\end{enumerate}

\subsection{Methodology}

\textbf{Baselines.} We compare \name{} with the following baselines. Note that neither vLLM nor SGLang natively supports prefix caching for Hybrid/SSM models, so we have extended both to support Hybrid models favorably.
\squishlist
    \item \textbf{Vanilla inference}: This baseline prefills all requests without doing prefix caching.
    \item \textbf{vLLM+}~\cite{vllm}: This baseline performs fine-grained checkpointing and caches a state for every token block. We use a token block size of 32, the largest size that vLLM supports~\cite{vllm_engine_args}, which favors vLLM+ by minimizing the number of low-utility SSM states admitted (Fig.~\ref{fig:cache_usage_breakdown}) while reducing memory fragmentation of KVs within a token block. %
    \item \textbf{SGLang+}~\cite{sglang}: While SGLang also uses a radix tree, it doesn't judiciously checkpoint states during admission. We enhance it by applying the same judicious admission policy as \name{}; however, the eviction policy remains LRU, which does not account for FLOP efficiency. %
\squishend

\textbf{Metrics.} 
The main metric we evaluate is token hit rate (\%), which represents the effectiveness of the cache and approximates the total compute saved (FLOP) well. We define token hit rate as the ratio of the number of tokens that skipped prefill over the total number of input tokens. We also evaluate different percentiles (P5, P50, and P95) of time to first token (TTFT, ms).
FLOP saved is a reasonable proxy for compute and latency savings because prefill is easily compute-bottlenecked. We do not evaluate downstream metrics (e.g., F1 score) as prefix reusing is exact and does not change the LLM output. %

\textbf{Workloads.} In our main results, we evaluate \name{} on two multi-turn conversational datasets: LMSys~\cite{lmsys} and ShareGPT~\cite{sharegpt} with different sequence lengths distributions. The LLM output sequences in LMSys are relatively long, often reaching thousands of tokens, whereas the LLM output sequences in ShareGPT are succinct and often take tens or hundreds of tokens. We also include an agentic workload: SWE-Agent~\cite{sweagent} on SWE-Bench~\cite{swebench}, a benchmark for evaluating LLM agents on real-world software issues collected from GitHub. We plot the sequence length distribution in Fig.~\ref{fig:input_len_distribution}.
These datasets contain multiple chat sessions, each with multiple rounds of requests. We vary the inter-session arrival time and inter-request arrival time to account for environment response time (e.g., human typing, interactions with the coding IDE) and queuing delay in the inference engine. 

\textbf{Models.} We use a 7B Hybrid model with \{4,24,28\} \{Attention,SSM,MLP\} layers in our main results.
For insights into our wins on TTFT, we use Jamba-1.5-Mini, an Attention-SSM Hybrid model with 12B active/52B total parameters, and serve it with state dimension 128 using the vLLM implementation on four A100-40GB GPUs.
We use FP16 precision in all experiments. The results apply to models of different sizes because the size of the prefix cache needs to be scaled accordingly.  %

\textbf{Setup.} Experiments were run on a p4d.24xlarge AWS instance with eight A100-40GB GPUs, 96 Intel(R) Xeon(R) Platinum 8275CL CPUs, and 1152GB of DDR4 RAM.

\subsection{End-to-End Results}

Fig.~\ref{fig:main_results} shows the token hit rate of vLLM+ and \name{} across different traces. The boxes show different quartiles of the data while the whiskers on the box plots show P5 and P95, allowing us to disregard extreme data and concentrate on the typical cases. With its judicious admission strategy, \name{} avoids wasteful caching decisions and optimizes the cache utilization, improving the hit rate by an average of 4.5$\times$, 7.3$\times$, and 34.4$\times$ for LMSys, ShareGPT, and SWEBench.

\begin{figure}[t]
    \centering
    \centering
    \includegraphics[width=0.9\linewidth]{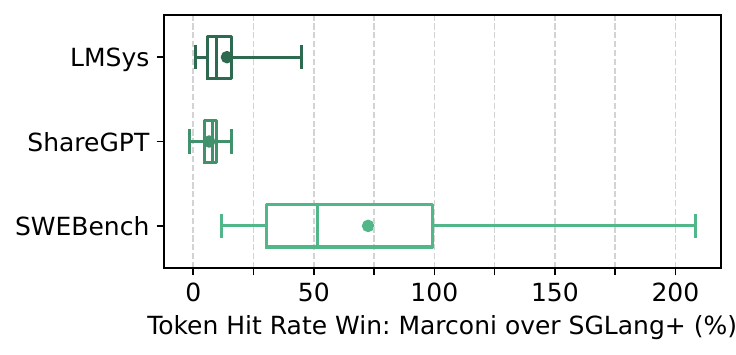}  %
    \vspace{-5pt}
    \caption{Comparison with SGLang+, which uses LRU as its eviction policy. \name{} balances recency and FLOPs efficiency, improving the token hit rate significantly, especially for workloads with longer context.}
    \label{fig:sglang_comparison}
    \vspace{-5pt}
\end{figure}

Fig.~\ref{fig:sglang_comparison} compares \name{} with SGLang+ to highlight the benefits of FLOP-aware eviction over LRU. The win is the most significant on SWEBench, with a P95 win of 219.7\%. The improvements on LMSys and ShareGPT are less pronounced, reaching a P95 win of 45.6\% and 19.0\%. This can be attributed to the difference in workload characteristics: as shown in Fig.~\ref{fig:input_len_distribution}, SWEBench has the widest input sequence length distribution, ranging from hundreds of tokens to tens of thousands of tokens. LMSys has a narrower distribution, with most sequences under 10K tokens, while ShareGPT predominantly features sequences under 2K tokens. The longer the sequences, the more critical FLOP efficiency becomes in eviction decisions. For workloads dominated by shorter sequences, suboptimal eviction choices in terms of FLOP efficiency have a smaller impact on \name{}'s performance gains.

\begin{figure}[t]
    \centering
    \centering
    \includegraphics[width=\linewidth]{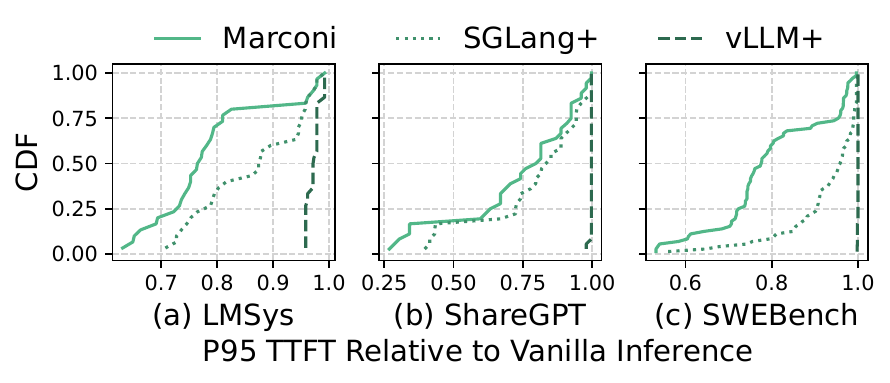}  %
    \vspace{-20pt}
    \caption{Distribution of P95 TTFT relative to no prefix caching.}
    \label{fig:ttft_total_distribution}
    \vspace{-10pt}
\end{figure}

Fig.~\ref{fig:ttft_total_distribution} shows the distribution of P95 TTFT relative to no prefix caching across different traces. \name{}'s token hit rate translates to TTFT reduction, reducing the P95 TTFT by up to 36.9\%, 73.2\%, and 46.8\% (281.4 ms, 106.3 ms, and 617.0 ms) compared to vanilla inference without prefix caching.
Compared to vLLM+, \name{} delivers up to 36.1\%, 71.1\%, and 46.8\% (275.4 ms, 103.3 ms, and 617.0 ms) larger P95 TTFT reductions; these numbers are 17.2\%, 12.8\%, and 24.7\% (131.1 ms, 18.5 ms, and 325.7 ms) when compared to SGLang+.

\subsection{Fine-Grained Analysis of FLOP-Aware Eviction}

\begin{figure}[t]
    \centering
    \begin{subfigure}{\linewidth}
        \centering
        \includegraphics[width=\linewidth]{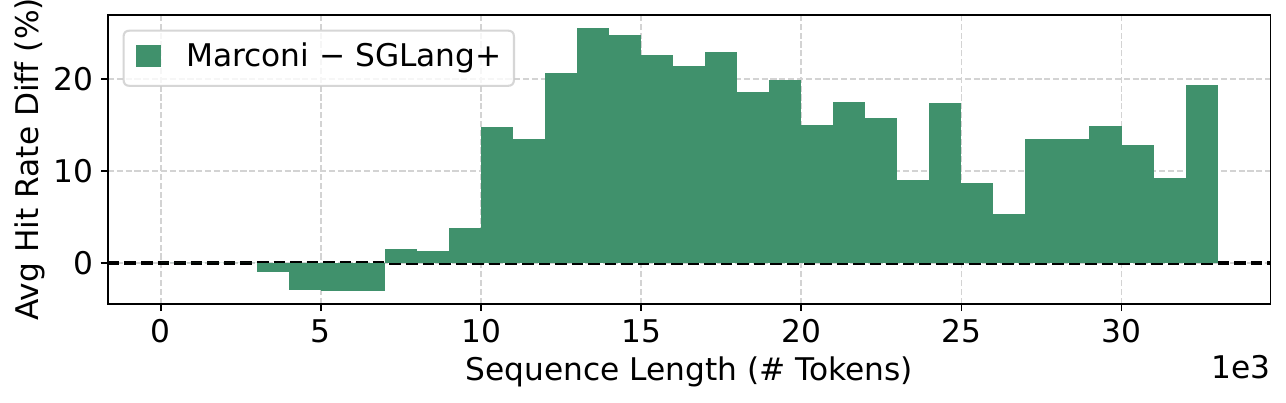}  %
        \caption{Difference between the average token hit rate of requests in \name{} and SGLang+, binned by the input sequence length.}
        \label{fig:token_hit_rate_comparison}
    \end{subfigure}
    \hfill
    \begin{subfigure}{\linewidth}
        \centering
        \includegraphics[width=\linewidth]{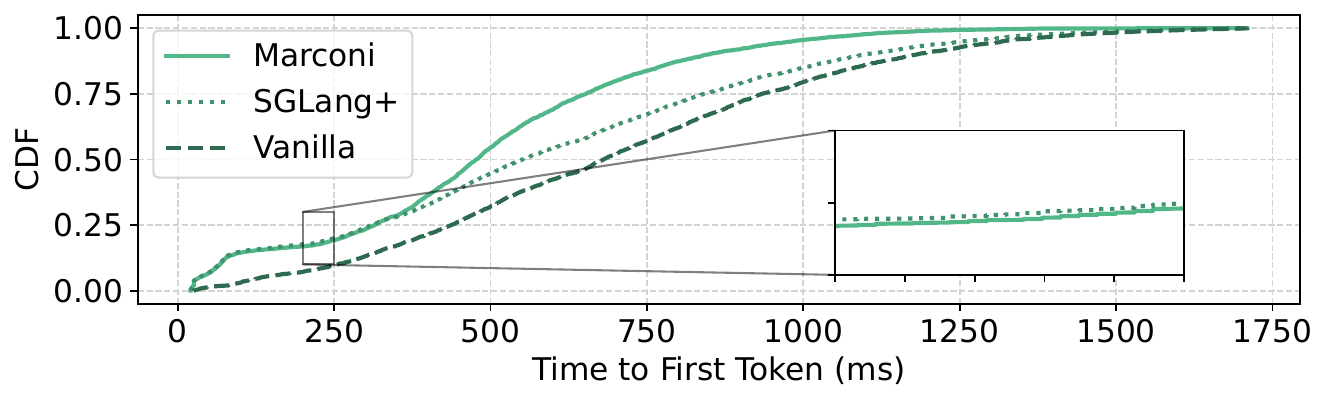}
        \caption{TTFT distribution of all requests in the trace.}
        \label{fig:ttft_distribution}
    \end{subfigure}
    \vspace{-15pt}
    \caption{Fine-grained analysis of FLOP-aware eviction. \name{} achieves a higher hit rate for longer sequences while sacrificing the hit rate for some shorter sequences (a), although the degradation in TTFT for shorter sequences is minimal (b).}
    \label{fig:fine_grained_analysis}
    \vspace{-5pt}
\end{figure}

To understand the performance improvement of FLOP-aware eviction over LRU, we compare the caching decisions 
of \name{} with SGLang+ using a request arrival trace from SWEBench. On this trace, SGLang+ achieves a 16.4\% overall token hit rate, while \name{} achieves a significantly higher hit rate of 32.7\%, an improvement of 99.4\%. 

In Fig.~\ref{fig:token_hit_rate_comparison}, we categorize the requests by sequence length and plot the difference in average hit rate between SGLang+ and \name{}. \name{} shows a lower hit rate (up to -3.0\%) for sequences with $<$7K tokens, while for sequences with $>$7K tokens, it outperforms SGLang+ with a hit rate improvement of up to 25.5\%. This is due to \name{}’s FLOP-aware approach, which prioritizes caching entries with higher FLOP efficiency under contention. Since longer sequences cost more FLOP, \name{} demonstrates an overall improvement of 90.3\% in FLOP saved compared to SGLang+.

Fig.~\ref{fig:ttft_distribution} shows the impact of FLOP-aware eviction on TTFT distribution. Due to \name{}'s lower hit rate for shorter sequences, it suffers a slight increase in TTFT at lower percentiles: \name{}'s P5 TTFT is 6.3\% worse than in SGLang+. However, because Hybrid models prefill short sequences quickly, the absolute latency reduction is minimal (2.1 ms, see the magnified area in Fig.~\ref{fig:ttft_distribution}). This trade-off allows \name{} to achieve a lower TTFT at higher percentiles, reducing P50 and P95 TTFT by 13.4\% and 22.0\% (74.2 ms and 274.9 ms), respectively.

\subsection{Microbenchmarks and Ablation Studies}

In these microbenchmarks, we use different representative traces to dissect \name{}'s performance improvements.

\begin{figure}[t]
    \centering
    \centering
    \includegraphics[width=0.75\linewidth]{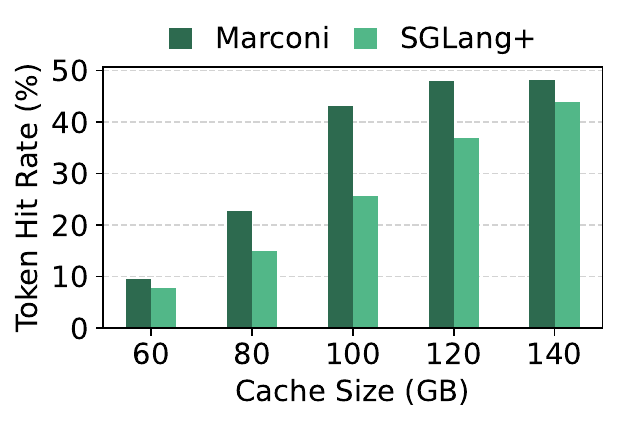}  %
    \vspace{-12pt}
    \caption{Impact of contention on FLOP-aware eviction's benefits. \name{} achieves the biggest win when the cache contention is moderate and judicious eviction decisions matter the most.}
    \vspace{-5pt}
    \label{fig:micro_contention}
\end{figure}

\textbf{Impact of cache contention.} In Fig.~\ref{fig:micro_contention}, we analyze how cache contention affects \name{}'s benefits. We vary the cache size from 60 GB (high contention) to 140 GB (low contention). 
Across the five cache sizes, \name{} achieves token hit rate improvements of 24.3\%, 51.5\%, 68.3\%, 30.0\%, and 10.0\% over SGLang+, respectively. The most significant performance gains occur under moderate contention, where eviction decisions are critical. In high contention scenarios, limited cache capacity prevents caching many useful prefixes (resulting in a token hit rate of less than 10\%). Conversely, in low contention scenarios, the cache has sufficient space to store a larger number of prefixes, and thus FLOP-unaware eviction decisions have a smaller impact on the hit rate.

\begin{figure}[t]
    \centering
    \begin{subfigure}[c]{0.52\linewidth}
        \centering
        \includegraphics[width=\linewidth]{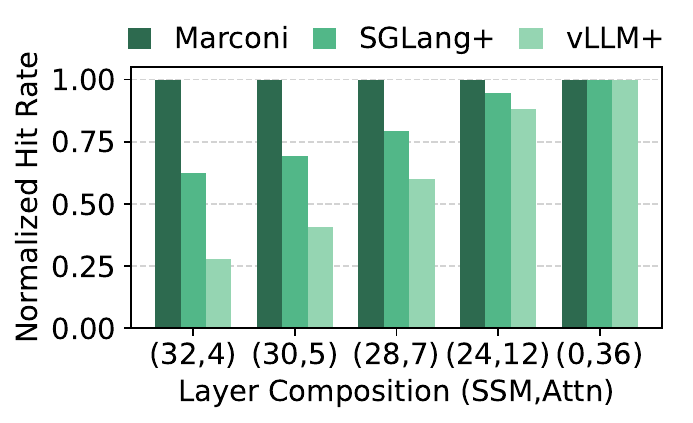}  %
        \caption{Varying layer compositions.}
        \label{fig:microbenchmark_layer_composition}
    \end{subfigure}
    \hfill
    \begin{subfigure}[c]{0.46\linewidth}
        \centering
        \includegraphics[width=\linewidth]{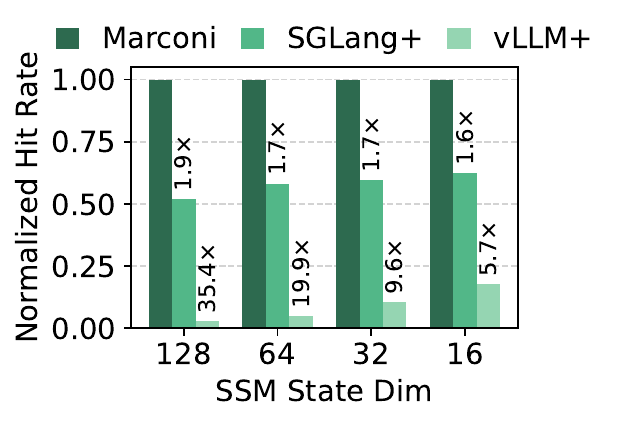}
        \caption{Varying state dimensions.}
        \label{fig:microbenchmark_state_dim}
    \end{subfigure}
    \caption{Impact of model architecture on \name{}'s performance. \name{} performs better for models with higher ratios of SSM layers and larger SSM state dimensions.}
    \label{fig:micro_model_impact}
    \vspace{-5pt}
\end{figure}

\textbf{Varying layer compositions.} As described in \S\ref{s:challenges}, the ratio of SSM layers directly affects the memory footprint of model states. In Fig.~\ref{fig:microbenchmark_layer_composition}, as we increase the Attention:SSM ratio from 1:2 to 1:4 to 1:8, \name{}'s token hit rate improvement over vLLM+ and SGLang+ increases from 13.5\% and 5.8\% to 66.6\% and 26.0\% to 2.6$\times$ and 59.7\%. When serving a pure Transformer, the three systems achieve the same performance. In order to get efficiency wins while preserving quality, the fundamental philosophy of Hybrid models is to use a \textit{majority} of subquadratic layers with efficient computational properties while mixing in a small number of Attention layers to ensure model quality~\cite{jamba,zamba,mamba_empicircal_study}. Therefore, we posit that \name{} will have better performance on most future Hybrid models.

\textbf{Varying SSM state dimensions.} As described in \S\ref{s:challenges}, the dimensionality of the recurrent SSM state directly affects the memory consumption of model states. Notably, the recent trend is for SSM state dimensionality to increase for better language modeling capability~\cite{mamba,mamba2}. In Fig.~\ref{fig:microbenchmark_state_dim}, as we increase the state dimension of a Hybrid model from 16 (Mamba1) to 128 (Mamba2), \name{}'s token hit rate improvement over vLLM+ grows from 5.7$\times$ to 35.4$\times$. A larger SSM state dimension increases the state sizes, exacerbating the issues in \S\ref{s:challenges} and making \name{}'s judicious admission more effective.

\begin{figure}[t]
    \centering
    \begin{subfigure}[c]{0.49\linewidth}
        \centering
        \includegraphics[width=\linewidth]{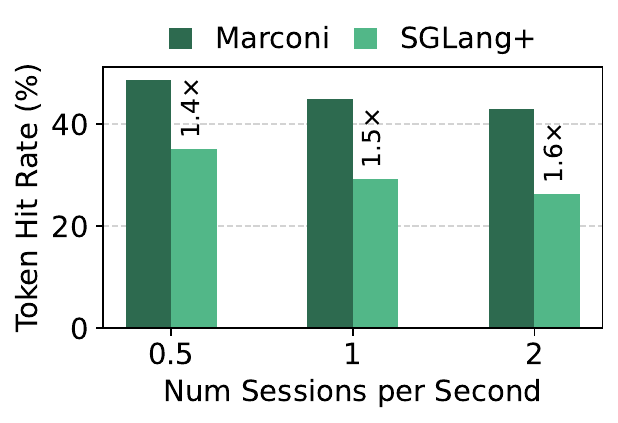}  %
        \caption{Varying session arrival rates.}
        \label{fig:micro_sps}
    \end{subfigure}
    \hfill
    \begin{subfigure}[c]{0.49\linewidth}
        \centering
        \includegraphics[width=\linewidth]{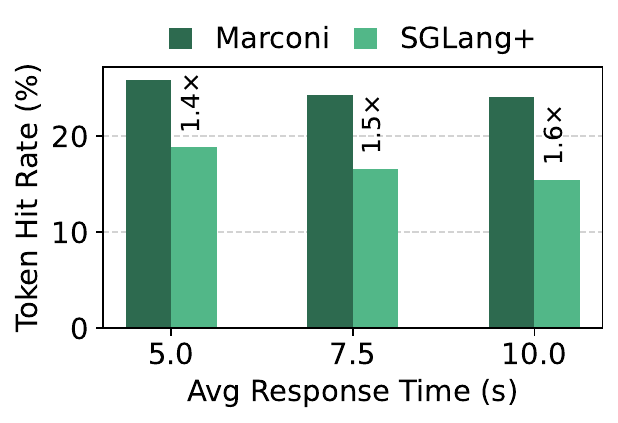}
        \caption{Varying request arrival rates.}
        \label{fig:micro_art}
    \end{subfigure}
    \caption{Impact of request arrival pattern on \name{}'s performance. Lower reusing opportunities, denoted by a higher session arrival rate (a) and longer time between requests within a session (b), reduce the token hit rate but improve \name{}'s relative win due to higher contention between requests.}
    \label{fig:micro_arrival}
    \vspace{-10pt}
\end{figure}

\textbf{Varying request arrival patterns.} Fig.~\ref{fig:micro_arrival} shows how request arrival patterns affect \name{}'s performance. 
As the average number of request (chat) sessions per second increases from 0.5 to 2, \name{}'s token hit rate decreases from 48.7\% to 43.0\%. Similarly, as the average response time between requests in a session increases from 5 s to 10 s, \name{}'s token hit rate decreases from 25.9\% to 24.1\% due to reduced effectiveness of prefix caching from more sessions sharing the fixed cache capacity and longer delays between prefix reuses in a session. However, \name{}'s relative improvement over SGLang+ grows from 1.4$\times$ to 1.6$\times$, thanks to increased contention between requests across sessions.

\section{Related Work}
\label{s:discussion}

\textbf{(Hybrid) recurrent subquadratic models.} 
There has been a resurgence of recurrent/linear models in the recent years: RWKV~\cite{rwkv}, RetNet~\cite{retnet}, GLA~\cite{gla}, Griffin~\cite{griffin}, RecurrentGemma~\cite{recurrentgemma}, xLSTM~\cite{xlstm}, Test-Time Training (TTT)~\cite{ttt}, DeltaNet~\cite{gated_deltanet, deltanet}, B'MOJO~\cite{bmojo}, MambaFormer~\cite{mambaformer}, Titans~\cite{titans}, Lightning Attention~\cite{minimax-01}, Hunyuan-TurboS~\cite{hunyuan}, Nemotron-H~\cite{nemotron_h}, etc. Importantly, although these models have different state updating rules~\cite{gla,deltanet}, they all update their (large) model states recurrently. As such, we only evaluate Mamba/SSMs, a representative architecture. The properties we summarize in~\S\ref{s:challenges} also apply to these recurrent layers (with slight variations on FLOP and state size), and \name{} can be extended to support prefix caching for all Hybrid models with recurrent layers. 

\textbf{Prefix caching.} Many recent systems have been proposed to capitalize on prefix reusing opportunities in LLM inference. InferCept~\cite{infercept} optimizes for KVs reusing in multi-turn chat scenarios. CachedAttention~\cite{cachedattention} and Pensieve~\cite{pensieve} maintain hierarchical caches to leverage memory/storage mediums. Preble~\cite{preble} performs cluster-level stateful caching that routes requests to the GPU with the longest prefix. PromptCache~\cite{promptcache} and CacheBlend~\cite{cacheblend} reuse KVs but do approximations that lead to accuracy drops. All past work admits all tokens' KVs, whereas \name{}'s philosophy of judicious admission is fundamentally different to handle SSM state entries' sparsity. Moreover, most prior work uses LRU for eviction and none of them factored in FLOP efficiency.

\textbf{Other LLM inference optimizations.}
Continuous batching is easier to realize for SSM layers because the tensors in each step of decoding don't have a sequence dimension and can easily be batched even if the sequences have different lengths, unlike in Transformers where selective batching is needed for applying batching only to certain operations~\cite{orca}.
The SSM states of Hybrid models don't require paged memory management~\cite{vllm} as they are fix-sized and don't grow and shrink like KVs. However, the KVs of Attention layers still need to be managed by paging.
Chunked prefill~\cite{sarathi} for Hybrid models requires specialized kernels under development in many serving frameworks~\cite{vllm}. Hybrid models benefit from layer-specific optimizations like FlashAttention~\cite{flashattention, flashattention2, flashattention3}.

\section{Conclusion}
\label{s:conclusion}

This paper proposes \name{}, the first prefix caching system designed to accommodate the unique characteristics of Hybrid models. \name{} proposes novel and judicious admission and eviction policies, achieving up to 34.4$\times$ higher token hit rates (71.1\% or 617 ms lower TTFT) over extended versions of state-of-the-art systems.

\section*{Acknowledgements}

We thank Yinwei Dai, Neil Agarwal, Mike Wong, and many fellow interns at AWS for their helpful discussions and feedback.

\nocite{langley00}

\bibliography{references}
\bibliographystyle{mlsys2024}

\appendix

\begin{table*}[h]
\centering
\begin{tabular}{llll}
\toprule
\multicolumn{1}{c}{} & \textbf{Attention} & \textbf{MLP} & \textbf{SSM}     \\
\midrule
\textbf{FLOPs per layer} & $8L D^2 + 4L^2 D$ & $16L D^2$ & $12L D^2 + 16LDN + 10L$      \\
\textbf{State size per layer (bytes)} & $4LD$ & N/A & $2DN$      \\
\midrule
\textbf{FLOPs saved per byte} & $L+2D$ & N/A & $L \cdot (6D/N + 8 + 5/DN)$ \\
\midrule
\textbf{FLOPs saved per byte (7B model)} & $L+8192$ & N/A & $200L$ \\
\bottomrule
\end{tabular}
\caption{The FLOP efficiency of different layer types. As seen in the last two rows, the FLOP efficiency of SSM layers scales much more steeply compared to Attention layers. SSMs' state sizes are orders of magnitude ($N/2=64$ in this 7B Hybrid model, where $D=4096$ and $N=128$) bigger than the KVs of a single token.}
\label{tab:flops_breakdown}
\end{table*}

\begin{table}[h]
\centering
\begin{tabular}{ll}
\toprule
\multicolumn{1}{c}{\textbf{Notation}} & \textbf{Description} \\
\midrule
L & Sequence length \\
D & Model dimension or $d\_model$ \\
N & State/feature dimension or $d\_state$ \\
\bottomrule
\end{tabular}
\caption{Glossary of notation and terminology.}
\label{tab:glossary}
\end{table}

\FloatBarrier

\section{Appendix}
\label{s:appendix}
\subsection{FLOP Efficiency Analysis}
In this section, we detail the math used for FLOP efficiency calculation. We list the notations used in Tab.~\ref{tab:glossary}.

\textbf{Memory footprint.} Assuming inference is performed in FP16, the memory size of the KVs in an Attention layer is calculated as 2 (K and V) $\cdot$ L $\cdot$ D $\cdot$ 2 bytes/parameter. In comparison, the memory size of an SSM layer's state is D $\cdot$ N $\cdot$ 2 bytes/parameter. Additionally, each SSM layer includes a \texttt{conv\_1d} block with a state size of \texttt{in\_channels} $\cdot$ \texttt{conv\_kernel} $\cdot$ 2 bytes/parameter. Since the \texttt{conv\_1d} states account for only a small fraction (6.1\% for the 7B Hybrid model used throughout the paper) of the total state size, we omit them from Tab.~\ref{tab:flops_breakdown} for simplicity, but they are included in all experiments in the main paper.

\begin{figure}[t]
    \centering
    \includegraphics[width=1.0\columnwidth]{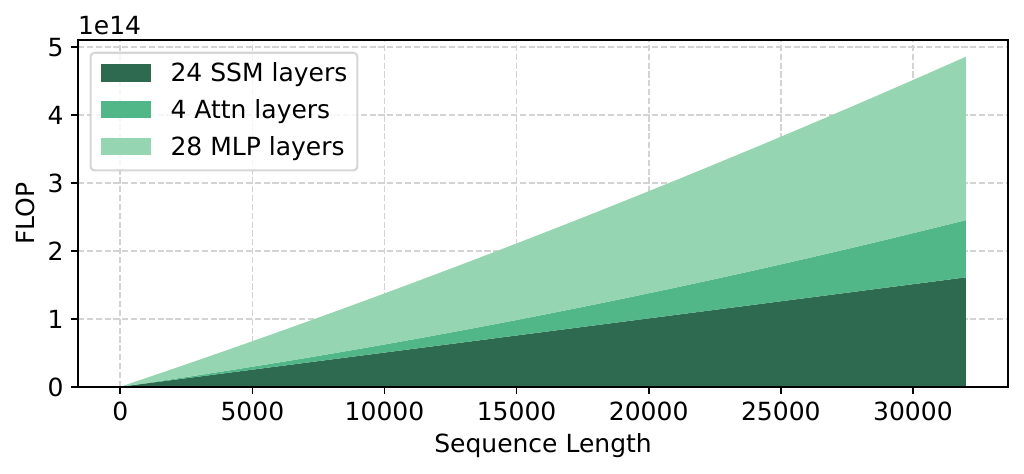}
    \caption{FLOP breakdown of a 7B Hybrid model.}
    \label{fig:flop_breakdown}
\end{figure}

\textbf{FLOP efficiency of models with different layer compositions.} Fig.\ref{fig:flop_breakdown} shows the FLOP distribution by layer type in a 7B hybrid model. Attention layers contribute fewer FLOPs for short sequences than SSMs and MLPs. However, as their quadratic computational complexity kicks in with longer sequences, they consume a significant portion of total FLOPs, even though they make up only 7.1\% of the model’s layers. Tab.\ref{tab:flops_breakdown} provides the FLOP breakdown by layer for a 7B hybrid model, showing that the FLOP efficiency of SSM layers scales more sharply compared to Attention layers.

\section{Artifact Appendix}

\subsection{Abstract}

In this artifact, we describe how to reproduce all experiments in this paper.
The evaluation utilizes request arrival traces from LMSys, ShareGPT, and SWEBench, all tokenized using the \texttt{meta-llama/Llama-2-7b-hf} tokenizer for consistency.
Key experiments involve running vLLM+, eviction policies V1 (SGLang+), V2 (Marconi), and V3 (offline-optimal, static-$\alpha$ oracle policy -- not included in the paper) across various dataset/arrival rate/cache size combinations.
This artifact supports easy customizations, including implementing additional eviction policies and evaluating new datasets and model configurations. For a detailed version of the artifact appendix, please see \url{https://github.com/ruipeterpan/marconi/blob/main/artifact_evaluation.md}.

\subsection{Artifact check-list (meta-information)}

{\small
\begin{itemize}
  \item {\bf Program: Python}
  \item {\bf Disk space required: $\sim$20 GB}
  \item {\bf Time needed to prepare workflow: $\sim$1 hour}
  \item {\bf Time needed to complete experiments: $\sim$12 hours}
  \item {\bf Publicly available: Yes}
  \item {\bf Code licenses: CC BY-NC}
  \item {\bf Archived DOI: \url{10.5281/zenodo.14970139}}
\end{itemize}

\subsection{Description}

\subsubsection{How delivered}

The artifact may be downloaded from zenodo at \url{https://zenodo.org/records/14970139} or cloned from the GitHub repository at \url{https://github.com/ruipeterpan/marconi}. The request arrival traces (with token-level information) can be downloaded from Google Drive by following the documentation below.

\subsubsection{Hardware dependencies}

We have tested Marconi on Cloudlab~\cite{cloudlab} nodes with Ubuntu 22.04 and Python 3.11.9. 
Due to the request arrival traces containing token IDs for each request, they require a total disk size of $\sim$7 GB to house.

\subsubsection{Software dependencies}

We have prepared a conda environment file that lists all dependencies and their versions.

\subsubsection{Datasets}

To ensure consistency when evaluating the same trace on different model architectures, we use the same tokenizer to produce the request arrival traces. For convenience, we compress the processed traces ($\sim$700M/6.3G pre/post compression) 
and host them on Google Drive.

\subsection{Installation}

The software dependencies can be installed via \texttt{conda} using the environment configuration file provided in the artifact. The traces can be downloaded via this \href{https://drive.google.com/file/d/1D8f68sBWJHyCfJZdEYCBK2M0iHmSDE6M/view?usp=sharing}{link}.

\subsection{Experiment workflow}

The sweep of all experiments (combinations of different cache sizes and request arrival patterns on different datasets) can be done by running \texttt{bash run\_all\_experiments.sh}, which invokes \texttt{policy\_exploration.py} and does the following for each experiment configuration (dataset/arrival rate/cache size combination):

\squishlist
    \item Runs vLLM+
    \item Runs eviction policy V1, which represents SGLang+
    \item Runs eviction policy V2, which represents Marconi
    \item Runs eviction policy V3, which represents an offline-optimal, static-\(\alpha\) oracle policy (the results weren't included in the paper). This policy sweeps over possible values of \(\alpha\) and selects the one that maximizes the hit rate
\squishend

Running \texttt{bash run\_all\_experiments.sh} creates three log files in \texttt{/logs}: \texttt{lmsys.txt}, \texttt{sharegpt.txt}, and \texttt{swebench.txt}. Each file contains the output log of all evaluations on the sweep of configurations for this dataset. Once the log files have been generated, plotting scripts can be run to analyze and plot the results.

\subsection{Evaluation and expected result}

All plotting scripts are under \texttt{/plotting} and can be run once the sweep of experiments on all configurations has finished:
\squishlist
    \item Fig. 7: \texttt{token\_hit\_rate.py}
    \item Fig. 8: \texttt{sglang\_comparison.py}
    \item Fig. 9: \texttt{ttft.py}
    \item Fig. 10: \texttt{fine\_grained\_analysis.py}
    \item Fig. 11: \texttt{microbenchmark\_contention.py}
    \item Fig. 12a: \texttt{microbenchmark\_layer\_composition.py}
    \item Fig. 12b: \texttt{microbenchmark\_dstate.py}
    \item Fig. 13a and 13b: \texttt{microbenchmark\_arrivalrate.py}
\squishend

\subsection{Experiment customization}

\squishlist
    \item Additional eviction policies can be easily implemented in \texttt{radix\_cache\_hybrid.py} by adding a new \texttt{evict\_policy\_version}
    \item Additional model configurations can be applied by editing the default configurations (NVIDIA's Attention-Mamba2 7B Hybrid model) in \texttt{policy\_exploration.py}
    \item Additional datasets can be evaluated by producing compatible traces as the ones generated by \texttt{/utils/generate\_trace.py}
\squishend

\subsection{Methodology}

Submission, reviewing, and badging methodology:

\squishlist
  \item \url{http://cTuning.org/ae/submission-20190109.html}
  \item \url{http://cTuning.org/ae/reviewing-20190109.html}
  \item \url{https://www.acm.org/publications/policies/artifact-review-badging}
\squishend

\end{document}